\documentstyle [12pt,a4wide]{article}
\def\ftilde{\widetilde F}
\def\htilde{\widetilde {\cal H}}

\def\dwf{W_F}

\def\dwhsq{W_H^2}

\def\dwfsq{W_ F^2}

\def\diffsigma{{\rm Diff}\Sigma}

\def\ldiffsigma{{\rm LDiff}\Sigma}

\def\gabd{\gamma_{\alpha\beta}}
\def\gabu{\gamma^{\alpha\beta}}
\def\ham{{\cal H}_{\perp}}
\def\hom{{\cal H}_{\perp ,i}}

\def\mom{{\cal H}_i }
\def\momj{{\cal H}_j }
\def\hamphi{{\cal H}_{\perp}^{\phi}}
\def\momphi{{\cal H}_i^{\phi} }
\def\momjphi{{\cal H}_j^{\phi} }
\def\hamtot{{\cal H}_{\perp}^T}
\def\momtot{{\cal H}_i^T }

\def\xxp{(x\leftrightarrow x')}
\def\xxpp{(ix\leftrightarrow jx')}
\def\ltilde{\lambda}
\def\mtilde{\mu}
\def\rtilde{R}

\def\mhf{\mu\bigl[\widetilde {\cal H},\widetilde F\bigr]}
\def\lhf{\lambda\bigl[\widetilde {\cal H},\widetilde F\bigr]}

\begin{document}  

\thispagestyle{empty}
\setcounter{page}{1}
\begin{titlepage}

\begin{flushright}
Imperial/TP/95--96/21.
\end{flushright}
\vskip 1cm
\begin{center}
{\large{\bf  Action functionals of single scalar fields and arbitrary--weight gravitational constraints that generate
a genuine Lie algebra.}}
\vskip 2cm
{ I.~Kouletsis\footnote{email: tpmsc17@ic.ac.uk. }
\vskip 0.4cm {\it Theoretical Physics Group,\\
Blackett Laboratory\\
Imperial College of Science, Technology \& Medicine\\
London  SW7 2BZ, U.K.}}
\vskip 0.7cm
\end{center}

\vskip 2.4cm
\begin{abstract}
We discuss the issue initiated by Kucha\v{r} {\it et al}, of replacing the usual Hamiltonian constraint by
alternative combinations of the
gravitational constraints (scalar densities of arbitrary
weight), whose Poisson brackets strongly vanish and
cast the standard constraint--system for vacuum gravity into a form
that generates a true Lie algebra. It is
shown 
that any such combination---that satisfies certain reality conditions---may be derived from an action principle
involving a single scalar field and a single Lagrange multiplier with
a non--derivative coupling to gravity.

\end{abstract}
\end{titlepage}

\renewcommand{\theequation}{\thesection.\arabic{equation}}
\let\ssection=\section
\renewcommand{\section}{\setcounter{equation}{0}\ssection}

\section{Introduction.}

The canonical formulation of general relativity is based on the
requirement that the spacetime manifold $\cal M$ is diffeomorphic with
$\Sigma\times R$. The real line $R$ plays the r\^ole of a global time
while the three--space $\Sigma$---which is assumed to be compact for
simplicity---represents physical space. The foliation map from
$\Sigma\times R$ to $\cal M$ induces an one--parameter family of embeddings from $\Sigma$ to $\cal M$ that are spacelike with respect to the four--metric on $\cal M$.

As a result, when the four metric (written as
$\gamma_{\alpha\beta}(X)$ in local coordinates) is pulled--back to
$\Sigma\times R$, a symmetric tensor field is induced on $\Sigma$ (the
spatial metric with local components $g_{ij}(x)$) that has signature
(1,1,1) and is positive definite. The other four independent components of the pull--back are related to the lapse function and the shift vector which determine how the hypersurface $\Sigma$ is displaced in spacetime under an infinitesimal change of global time.

The canonical version of general relativity, for example see
\cite{Isham}, is derived by decomposing the usual Einstein Lagrangian
density with respect to the spatial metric, the shift vector and the
lapse function, and then performing a Legendre transformation to replace the time derivative of the three--metric with its conjugate momentum $p^{ij}(x)$. In doing so, the lapse function and the shift vector become non--dynamical Lagrange multipliers, enforcing on the canonical variables ($g_{ij}(x)$, $p^{ij}(x)$) the Hamiltonian and momentum constraints\footnote{
Spacetime points are denoted by $X$ and spatial points by $x$. Greek
letters $\alpha,\beta,...$ are used as spacetime indices and run from
0 to 3 while  Latin characters $i,j,...$ are used as spatial indices
and run from 1 to 3. The notation for functions, functionals, ...e.t.c., is not consistent but rather obeys the requirement of maximum
notational simplicity under each particular situation. }\cite{{ADM},{KuDGR}},
\begin{eqnarray}
\ham(x)&=&G_{ijkl}(x)p^{ij}(x)p^{kl}(x)-g^{1\over2}(x)\ ^{(3)}R[g_{ij}(x)]\nonumber\\
    & &G_{ijkl}(x)={1\over2}g^{-{1\over2}}(x)\biggl(
g_{ik}(x)g_{jl}(x)+g_{il}(x)g_{jk}(x)-g_{ij}(x)g_{kl}(x)\biggr)\nonumber\\
\mom(x)&=&-2D_jp^j_i(x),
\label{eq:constraint}
\end{eqnarray}
where $g(x)$ denotes the determinant of the spatial metric $g_{ij}(x)$. 

The constraints
\begin{equation}
\ham(x)=0=\mom(x)
\label{eq:con}
\end{equation}
satisfy the Dirac algebra
 \cite{Dirac}, 
\begin{eqnarray}
\{\ham (x), \ham (x')\}&=&g^{ij}(x)\mom(x) \delta_{,j}(x,x')-\xxp\label{eq:D1}
\\
\{\ham(x), \mom(x')\}&=&\hom(x) \delta (x,x')+
\ham(x) \delta_{,i}(x,x')\label{eq:D2}
\\
\{\mom(x),\momj (x')\}&=&\momj(x)\delta_{,i}(x,x')-\xxpp, \label{eq:D3}
\end {eqnarray}
and are first--class since the right hand side of equations
(\ref{eq:D1})--(\ref{eq:D3}) vanishes on the constraint surface
(\ref{eq:con}). The appearence of the $g^{ij}(x)$ factor in
(\ref{eq:D1}), however, implies that the algebra they generate is not
a genuine {\it Lie algebra}---a property that creates various problems in a possible quantum version of the theory.

While studying the coupling of gravity to dust, Brown and Kucha\v{r}
\cite{BrKu} came across a weight--two scalar combination of the gravitational constraints,  
\begin{equation}
G(x):=\ham ^2(x)-g^{ij}\mom(x) \momj(x),
\label{eq:G}
\end {equation}
that has strongly vanishing Poisson brackets with itself. The
combination was found equal to the square of the momentum conjugate to
the dust time and, as such, it had to be positive;
equivalently, the gravitational sector of the phase space of the
theory was
limited to certain appropriate regions.

If $G(x)$ replaces the usual Hamiltonian constraint to form an---at least
locally---equivalent new set of constraints for vacuum general relativity, 
\begin{equation}
G(x)=0=\mom(x),
\end{equation}
a genuine Lie algebra is created: 
\begin{eqnarray}
\{G(x),G(x')\}&=&0\,\label{eq:GG} \\
\{G(x), \mom (x')\}&=&G_{,i}(x)\delta (x,x')+2G(x)\delta_{,i}(x,x')\,
\label{eq:GH}\\
\{\mom (x), \momj (x')\}&=&\momj(x)\delta_{,i}(x,x')-\xxpp. \label{eq:HH} 
\end {eqnarray} 
It corresponds to the semidirect product of the Abelian algebra 
generated by $G(x)$, equation (\ref{eq:GG}), and the algebra of 
spatial diffeomorphisms 
$\ldiffsigma$ produced by $\mom(x)$, equation (\ref{eq:HH}). 
The Poisson bracket (\ref{eq:GH}) reflects the transformation 
of $G(x)$ as a weight--two
scalar density under $\diffsigma$.

A similar result was obtained by Kucha\v{r} and Romano \cite{KuRo}. They coupled gravity to a single massless scalar field and extracted the following weight--two scalar combination of the gravitational constraints:   
\begin{equation}
\Lambda_{\pm}(x):=g^{1\over2}(x)\biggl(-\ham (x)\pm\sqrt{G(x)}\biggr). \label{eq:L}
\end {equation} 
The combination was found to be equal to the square of the momentum
conjugate to the scalar field and has exactly the same Poisson brackets
as $G(x)$, given by equations (\ref{eq:GG})--(\ref{eq:HH}). If a proper choice
of sign for the square root is made\footnote{See the end of sections 2
and 3.}, $\Lambda_{\pm}(x)$ may be used to replace the Hamiltonian
constraint in a new locally equivalent system of constraints for pure gravity. Again, the theory allows only certain regions of
the gravitational phase space so that the quantity
inside the square root and the combination
${\Lambda}_{\pm}(x)$ itself be positive.
 
The question that arises \cite{KuRo} is whether some other couplings
of gravity to fields can also lead to combinations of the
gravitational constraints with strongly vanishing Poisson brackets,
and whether the multiplicity of such possible alternative combinations
conveys any general message about the structure of canonical general
relativity.

A significant advance was made recently by Markopoulou \cite{Mark}, who
constructed a unique nonlinear partial differential equation satisfied
by any scalar combination of arbitrary weight that obeys the abelian
algebra (\ref{eq:GG}). Because of the extensive overlap between
\cite{Mark} and the present paper, a brief review of Markopoulou's
results is presented. 

The basic observation was that any arbitrary--weight scalar density
(which in \cite{Mark} was collectively denoted by ${\cal W}(x)$) can be written as a general function of the simplest possible scalar combinations of the gravitational constraints and the determinant of the spatial metric\footnote{For partial differentials the notation 
$A_b(b):=\partial A(b)/\partial b$ is used. 
Very often in what follows, partial differentials with respect to 
$\htilde(x):=g^{-{1\over 2}}(x)\ham(x)$ and 
$\ftilde(x):=g^{-1}(x)g^{ij}(x)\mom(x)\momj(x)$ appear; 
they are denoted by 
$A_H(x):=\partial A[\htilde(x),\ftilde(x)]/\partial\htilde(x)$ 
and 
$A_F(x):=\partial A[\htilde(x),\ftilde(x)]/\partial\ftilde(x)$ 
respectively.}, 
\begin{equation}
{\cal W}_{\omega}[\htilde(x),\ftilde(x),g(x)]=g^{\omega\over 2}(x) W_{\omega}[\htilde(x),\ftilde(x)],
\label{eq:skata}
\end{equation}
on the assumption that ${\cal W}_{\omega}[\htilde(x),\ftilde(x)]$ is an {\it
ultralocal} \cite{IsKu} function of them.  
The two basic {\it weight--zero} combinations of the constraints are
defined by $\htilde(x) := g^{-{1\over 2}}(x)\ham(x)$ and $\ftilde(x)
:= g^{-1}(x) g^{ij}(x)\mom(x)\momj(x)$, while the parameter $\omega$
denotes the weight of the relevant scalar densities, and is now
also used as a subscript.

The requirement of strongly vanishing Poisson brackets was then
imposed on ${\cal W}_{\omega}(x)$, leading to an unexpectedly compact
and exactly solvable differential equation for the weightless part of
(\ref{eq:skata}) (the subscript $\omega$ has been dropped),
\begin{equation}
{ \omega \over 2} W(x) \dwf(x) = \ftilde(x) \dwfsq(x) - {1\over 4}\dwhsq(x),
\label{eq:WPDE}
\end {equation}
which must hold at every spatial point {\it x}. 
The uniqueness of equation (\ref{eq:WPDE}) is assured by the kind of assumption made on the form of ${\cal W}_{\omega}(x)$, which is the least restrective one. Its general solution is found to be
\begin{eqnarray}
W_{\omega}\bigl[\htilde, \ftilde, B(\alpha[\htilde,\ftilde])\bigr]
& = & \pm
\Biggl[\biggl(\htilde - {1\over 2}B'(\alpha[\htilde,\ftilde])
\biggr) +\sqrt
{\biggl(\htilde - {1\over 2} B'(\alpha[\htilde,\ftilde])
\biggr)-\ftilde} \ \Biggr]
^{\omega\over 2} \nonumber \\
& & \times \exp\Biggl(B(\alpha[\htilde,\ftilde])
+{\omega\over2}
{ {1\over2} B'(\alpha[\htilde,\ftilde]) \over 
\sqrt{\bigl(\htilde - {1\over 2}B'(\alpha[\htilde,\ftilde])
\bigr)-\ftilde}}\Biggr),
\label{eq:aasol}
\end{eqnarray}
where the {\it x}'s have been omitted, and $\alpha[\htilde,\ftilde]$ is
determined by algebraically solving the equation 
\begin{equation}
\alpha=-{\omega\over 4\sqrt{(\htilde-{1\over 2}
B'(\alpha))^2-\ftilde}}
\label{eq:asol}
\end{equation}
for a given choice of $B(\alpha)$. {\it Complex} solutions for
$W_{\omega}(x)$ can in general exist. 

Expressions (\ref{eq:aasol}) and (\ref{eq:asol})---which provide a family of self--commuting constraint
combinations parametrised by an
arbitrary function of one variable---were based on purely algebraic
considerations \cite{Mark} and,
consequently, their physical relevance is not clear. However, at least
for the special case of weight two, a lot of
insight into their origin could be gained if they were
shown to be related to some phenomenological physical system similar
to the ones discussed in
\cite{BrKu} and \cite{KuRo}. 
An attractive possibility---supported by the fact that both combinations $G(x)$ and $\Lambda_{\pm}(x)$ appear in
(\ref{eq:aasol}), (\ref{eq:asol}) as genuine subcases of the same weight--two
solution---would be for the action functionals for dust \cite{BrKu} and
for a massless scalar field \cite{KuRo} to both arise as different
versions of a wider, generalised action functional parametrised by an
arbitrary function of one variable.

An obvious objection to such a proposal comes from the fact that
the dust action \cite{BrKu} involved {\it four} scalar fields compared to
the {\it single} scalar field that was considered in \cite{KuRo}. A closer analysis
however suggests that, practically speaking, it was the
``one--term--factorisation'' property of the dust action that was
responsible for the succesful extraction of the $G(x)$, or, to put it
in another way, it was the fact that the dust action strongly
resembled the simpler action for a massless scalar
field. Indeed, it is straightforward to show that $G(x)$ can also be
derived from a variety of single--scalar--field actions which---although
perhaps not as convincingly interpretable as the one for dust---are
still equally effective. 

Motivated by this observation, one may use some
action functional of a single scalar field---having the two important properties
of being parametrised by an arbitrary function of one variable and 
reducing to the actions of \cite{BrKu} and \cite{KuRo} for certain
choices of this function---and
hope that, when coupled to
gravity, it will provide the general solution of equation (\ref{eq:WPDE})
for weight two (at least the subset of the general solution that can have real positive
values in some regions of the gravitational phase space, as
in the cases of \cite{BrKu} and \cite{KuRo}). In addition,
due to a remarkable feature of equation (\ref{eq:WPDE})---namely, if $W_{\omega}(x)$ is a
solution of weight $\omega$ then $W_{\omega'}(x)$ := ${W_{\omega}}^{{\omega'}\over{\omega}}(x)$ is 
a corresponding solution\footnote{Provided that both
$\omega$ and $\omega'$ are different from zero, so that the
algorithm is well--defined and invertible. In particular,
if $\omega$ equals zero, the left
hand side of (\ref{eq:WPDE}) vanishes and the equation
becomes a homogeneous one, i.e., a {\it different} equation---this
degenerate case is therefore excluded from the present discussion.} of weight ${\omega}'$---one expects that if
the weight--two procedure proves to be successful, may be naturally
extendable to arbitrary weight. 
The above argument actually works and is the key to the present analysis,
which is organised as follows:

In the introduction to section 2, we make some preliminary remarks about
differential equation (\ref{eq:WPDE}) and the existence of complex solutions
(\ref{eq:aasol}), (\ref{eq:asol}), as
well as some general comments about the present paper.
As claimed above, equation (\ref{eq:WPDE}) is for every weight directly related to a
generalised action functional of a single scalar field, parametrised by an arbitrary
function of one variable. This is presented in section 3. As it stands,
however, the equation does not make this connection clear and
therefore a suitable ansatz is used (in section 2) to convert the weight--zero part
of the scalar densities $W_{\omega}(x)$ into a form that is better suited
for our purposes.

The so--called $\omega$--ansatz (since there is one for each weight) expresses $W_{\omega}(x)$ in terms
of two weight--zero complex
scalar densities $\ltilde(x)$ and $\mtilde(x)$ which, as
$W_{\omega}(x)$, are {\it ultralocal} functions of the simplest possible scalar
constraint--combinations $\htilde(x)$ and $\ftilde(x)$\footnote{Although the
differential equation (\ref{eq:WPDE}) admits {\it complex--valued}
solutions that cannot be reconciled to the idea of the physical system
of section 3, it is chosen---for the sake of uniformity---not
to impose at this stage any reality--conditions on $W_{\omega}(x)$,
leaving the necessary adjustments for the physical relevance of these
solutions to be done in section 4. It should be mentioned here---for
more details see the introduction of section 2---that the terms
``complex'' and ``complex--valued'' do not imply each other.}.   
The ansatz essentially transforms the nonlinear arbitrary--weight
differential equation (\ref{eq:WPDE}) into four different pairs of
coupled  quasilinear partial differential equations for $\ltilde(x)$ and
$\mtilde(x)$. This transformed version of the ``$\omega$--equation'' has
the property of being {\it
weight--independent} with all the information about the relevant weight 
contained in the $\omega$--ansatz.

More precisely, any (weight--independent) solution of a particular
pair of equations for $\mu(x)$ and $\lambda(x)$
is mapped---through the (weight--dependent) $\omega$--ansatz---to a 
solution of the differential equation (\ref{eq:WPDE}) of the
corresponding weight\footnote{Excluding some trivial cases where the
partial derivatives of $W_{\omega}(x)$ {\it identically} vanish; see introduction of
section 2.}. A crucial feature is that the ansatz map, denoted by
$a_{\omega}$, is {\it onto} and it can be made {\it one--to--one} if certain
(weight--dependent) equivalence
classes of solutions in the domain set of each $a_{\omega}$ are defined.
This implies that any one of the
four ``linearised'' pairs of equations for $\ltilde(x)$ and $\mtilde(x)$ 
is exactly equivalent to the original nonlinear $\omega$--equation for
$W_{\omega}(x)$, through the appropriate $a_{\omega}$ map. The most 
symmetric pair of the four is then singled out and considered 
as the ``representative'' one. It is the pair of equations that is going to be related to the action principle.

In section 3, the relevant scalar--field action is introduced and
involves a single scalar field and two initially arbitrary
{\it real--valued}  functions 
$\ltilde(M(X))$ and $\mtilde(M(X))$ of a single (real or complex)
Lagrange multiplier M(X)\footnote{The inclusion of complex
multipliers is because they are {\it allowed} by expressions
(\ref{eq:aasol}), (\ref{eq:asol}) and because they do not spoil the
physical ``realizability''
of the scalar--field system, as it is properly explained in the
introduction to section 2}.
It is the simplest possible action--candidate that includes as genuine
subcases both actions of \cite{BrKu} and \cite{KuRo}---the former in
a single--field version---and possesses the required parametrisation by an
arbitrary function of one variable, after the elimination of the
non--dynamical multiplier.
The scalar--field action is coupled to the Einstein--Hilbert one, and
the Hamiltonian form of the total action is obtained by the usual
A.D.M. decomposition. It follows the general rule that is common to
all theories with a non--derivative coupling to gravity.

When the $\omega\over2$--power of the square of the weightless field momentum 
$(\pi/g^{1\over 2})(x)$---in the coupled Hamiltonian system---is solved
with respect to the gravitational variables alone,
it takes the form of the $\omega$--ansatz of section 2, 
with $\ltilde(x)$ and $\mtilde(x)$ replaced by the two functions of the
multiplier $\ltilde(M(x))$ and $\mtilde(M(x))$ (being now in canonical form). 
Furthermore, when the multiplier $M(x)$ and hence these two functions
are also solved
with respect to the gravitational variables,  $\ltilde(M(x))$ and
$\mtilde(M(x))$ become functionally dependent and 
are surprisingly shown to satisfy the ``representative'' pair of equations!
In other words,
$(\pi^{2}/g)^{\omega\over 2}(x)$ {\it is bound}
to produce solutions $W_{\omega}(x)$ of the corresponding $\omega$--equation, for
{\it every} initial choice of functions $\ltilde(M(x))$ and
$\mtilde(M(x))$\footnote{Apart from a few choices that lead either to a constraint
between $\htilde$ and $\ftilde$ (see end of section 3) or to solutions
for the multiplier $M(x)$ that are not consistent with the requirement
that $\ltilde(M(X))$ and $\mtilde(M(X))$ be real--valued. Both
these problems, however, are finally avoided in section 4.}. 

In section 4, the inverse procedure is analysed and it is shown that
{\it any} solution $W_{\omega}(x)$ that satisfies certain
conditions of reality---suggested by the qualitative predictions of the
preceding sections---can be derived from the scalar--field
action principle of section 3; under these conditions therefore, the latter provides the {\it general solution} of the $\omega$--equation. 
Considering the diverse origins
of the calculations in sections 2 and 3, this is a
striking result and suggests that the r\^ole of scalar fields in classical 
general relativity deserves to be investigated further. 

The connection  
between the $\ltilde's$, $\mtilde's$ and the notation\cite{Mark} used in the
explicit solutions (\ref{eq:aasol}) and (\ref{eq:asol}) is established in one of the appendices.

\section{An equivalent form  of the weight--$\omega$ differential  equation.}

\subsection{General remarks.}
A minimum prerequisite concerning the physical relevance in vacuum
gravity of the gravitational combinations that satisfy equation
(\ref{eq:WPDE}), is for them to create a constraint surface that is
locally 
identical to the usual one. Then, by replacing the Hamiltonian
constraint, they can form an {\it at least} locally equivalent system
of constraints, that has the additional property of generating a true Lie algebra.
However, by concentrating on the actual form of equation (\ref{eq:WPDE}), one
observes that in the particular case when the partial derivatives of
some solution $W_{\omega}$ with respect to either $\htilde$ or $\ftilde$ happen
to be {\it identically} zero, this solution is either a function of $\ftilde$ alone or a
constant. In both cases the solution $W_{\omega}$ does not depend on the
Hamiltonian constraint and all information concerning the latter is
lost. These special solutions of equation (\ref{eq:WPDE})---when the quantities $W_{{\omega}F}$, $W_{{\omega}H}$ and of course $W_{\omega}$ identically vanish---can therefore be safely ignored as inadequate for any physical application.

On the other hand, the existence of complex solutions of equation
(\ref{eq:WPDE}) is a feature that 
certainly deserves some attention. It is true that such solutions
cannot be easily reconciled to the idea of a physical
system---especially in the present discussion when, later, they are related to the
$\omega\over2$ power
 of the {\it square} of some conjugate field--momentum. However, a {\it complex}
combination of the gravitational constraints is not necessarily
{\it complex--valued}, as the simple example $i
\sqrt{{\htilde}^2-\ftilde}$ (that solves the $\omega$=1 equation) demonstrates in an appropriate region
of the phase space where $\ftilde>{\htilde}^2$. 
Indeed, the above complex combination is merely an equivalent way of writing
$\sqrt{\ftilde-{\htilde}^2}$ which---in the same region of the phase
space---is both {\it real} and {\it real--valued}.   

For the sake of uniformity, therefore, and in order to avoid using
adjectives like ``complex'' and ``complex--valued'' throughout this
section, it seems preferable to
accept any 
generality offered by the two
expressions (\ref{eq:aasol}) and
(\ref{eq:asol}) that solve the differential equation
(\ref{eq:WPDE}), and allow---at this stage---negative and even complex--valued 
combinations of the gravitational constraints\footnote{As, for
example, the solutions $-({\htilde}^2-\ftilde)^2$ and
$i({\htilde}^2-\ftilde)$ that solve the $\omega=4$ and $\omega=2$ equations, respectively.}. The 
necessary amendments for the physical relevance of the combinations
is
reserved for section 4.  

\subsection{The $\omega$--ansatz and the ``representative'' equation.}
As already mentioned, equation (\ref{eq:WPDE}) will be related in
section 3 to a quite
general action principle involving a single scalar field and two arbitrary
functions of a Lagrange multiplier. For this reason, 
an appropriate $\omega$--ansatz is introduced---one for each weight---that transforms the corresponding
$\omega$--equation (\ref{eq:WPDE}) into a form that
is easily connected to this action principle. The ansatz is:
\begin{equation}
W_{\omega}[\htilde,\ftilde]=\ltilde^{\omega\over 2}[\htilde,\ftilde]
\Biggl(\htilde-\mtilde[\htilde,\ftilde]+
\sqrt{\bigl(\htilde-\mtilde[\htilde,\ftilde]\bigl)^2-\ftilde}
\Biggr)^{\omega\over 2},\label{eq:ansatz}
\end{equation}
where $W_{\omega}[\htilde,\ftilde]$, $\mhf$
and $\lhf$ are ultralocal functions of $\htilde$ and $\ftilde$,
and are allowed---in general---to be complex--valued.

Both signs for the square root in equation (\ref{eq:ansatz})
are permitted; this is not denoted
by a $\pm$ sign just for notational simplicity.
Note that any solution, say ${\mit v}_{\omega}[\htilde,\ftilde]$,
 of the differential
equation (\ref{eq:WPDE})---obtained by any means---can always be brought
into the form of the ansatz by choosing 
\begin{equation}
\lambda^{\omega\over 2}[\htilde,\ftilde]={\mit v}_{\omega}[\htilde,\ftilde]
\Biggl(\htilde-\mtilde[\htilde,\ftilde]+
\sqrt{\bigl(\htilde-\mtilde[\htilde,\ftilde]\bigl)^2-\ftilde}
\Biggr)^{-{\omega\over 2}}
\end{equation}
while keeping $\mu[\htilde,\ftilde]$ arbitrary. There is no loss of
generality, therefore, in writing any solution ${\mit v}_{\omega}$ of
(\ref{eq:WPDE}) in the form of equation (\ref{eq:ansatz}). 

It must be said here that, strictly speaking, our arguments in all that follows must be restricted
to regions of the $\htilde$, $\ftilde$ plane where every function
of the constraints
as well as its partial derivatives are well--defined.
This implies the necessity---at least for most of the cases---to perform all calculations away from the usual
constraint surface of vacuum general relativity, and return to
it only after the completion of the calculations. It also clearly emphasizes
the difficulty in assigning any proper meaning to the differential equation
(\ref{eq:WPDE}) without invoking some medium to which gravity is
presumably coupled. Having mentioned that, the square root in (\ref{eq:ansatz}) is
defined as
\begin{equation}
\rtilde[\htilde,\ftilde]:=\sqrt{\biggl(\htilde-\mtilde[\htilde,\ftilde]\biggl)^2-\ftilde},
\label{eq:root}
\end{equation}
and the two partial derivatives of
$W_{\omega}[\htilde,\ftilde]$ in terms of  $\lhf$,
$\mhf$ and $R[\htilde,\ftilde]$ 
are obtained (the square bracket notation will be omitted from now on):
\begin{eqnarray}
W_{{\omega}H}&=&
{\omega\over 2}\ltilde^{\omega\over 2}(\htilde-\mtilde+\rtilde)^{\omega\over 2}\biggl(
{1\over\ltilde}\lambda_H-
{1\over\rtilde}\mu_H+{1\over\rtilde} \biggr)\nonumber\\
W_{{\omega}F}&=&
{\omega\over 2}\ltilde^{\omega\over 2}(\htilde-\mtilde+\rtilde)^{\omega\over 2}\Bigl(
{1\over\ltilde}\lambda_F-
{1\over\rtilde}\mu_F-{1\over{2\rtilde(\htilde-\mtilde+\rtilde)}}\Bigr).
\label{eq:vf}
\end{eqnarray}

When the ansatz (\ref{eq:ansatz}) and the derivatives
 (\ref{eq:vf}) are substituted into the differential equation (\ref{eq:WPDE}), the latter becomes:
\begin{equation}
\biggl({1\over\ltilde}\lambda_F-
{1\over\rtilde}\mu_F\biggr)
\Biggl[-\ftilde \biggl(
{1\over\ltilde}\lambda_F-
{1\over\rtilde}\mu_F\biggr)
+{\htilde-\mtilde\over\rtilde}\Biggr]+
{1\over4}\biggl(
{1\over\ltilde}\lambda_H-
{1\over\rtilde}\mu_H\biggr)
\Biggl[\biggl(
{1\over\ltilde}\lambda_H-
{1\over\rtilde}\mu_H\Bigr)
+{2\over\rtilde}\Biggr]=0, \label{eq:square}
\end{equation}
where, quite noticeably, the arbitrary weight $\omega$ is no longer
present.
 
It can be seen, by inspection, that there exist four obvious solutions of (\ref{eq:square}), corresponding to
 four different pairs of coupled quasilinear
equations for the functions $\mtilde$ and $\ltilde$. These quasilinear equations are written explicitly below:

\begin{eqnarray}
{1\over\ltilde}\lambda_F-
{1\over\rtilde}\mu_F=0&
{\rm and}& 
{1\over\ltilde}\lambda_H-
{1\over\rtilde}\mu_H=0,
\label{eq:a}\\
{1\over\ltilde}\lambda_F-
{1\over\rtilde} \mu_F=0  &
{\rm and}&     
{1\over\ltilde}\lambda_H-
{1\over\rtilde}\mu_H=-{2\over\rtilde},
\label{eq:b}\\
{1\over\ltilde}\lambda_F-
{1\over\rtilde}\mu_F
={\htilde-\mtilde\over\rtilde}
      &
{\rm and}&     
{1\over\ltilde}\lambda_H-
{1\over\rtilde}\mu_H=0,
\label{eq:c}\\
{1\over\ltilde}\lambda_F-
{1\over\rtilde}\mu_F
={\htilde-\mtilde\over\rtilde}    &
{\rm and}&      
{1\over\ltilde}\lambda_H-
{1\over\rtilde}\mu_H=-{2\over\rtilde}.
\label{eq:d}
\end{eqnarray}

Given a $\mtilde$, any of the above pairs of
equations can be solved for the corresponding $\ltilde$---provided that the
system of the two partial equations for $\ltilde$ is not
contradictory---and then the $\omega$--ansatz (\ref{eq:ansatz}) can be used to
obtain solutions $W_{\omega}$ of equation (\ref{eq:WPDE}) of the
corresponding weight.
An equivalent procedure can be followed if, instead of $\mu$, a $\lambda$ is initially chosen. In other words, each of the above four 
pairs of equations (\ref{eq:a})--(\ref{eq:d}) provides---for each
weight---a family 
of solutions of the corresponding differential equation, that is 
parametrized by an arbitrary function
$\mtilde$ or $\ltilde$ and is subject to the condition that the
relevant pair of equations from the set (\ref{eq:a})--(\ref{eq:d}) 
is not self--contradictory.

Furthermore, the above pairs are all
{\it equivalent} in the sense that, for each weight, they all lead to exactly the {\it
same} family of solutions of the differential equation (\ref{eq:WPDE}). More precisely, for
every solution $W_{\omega}$ of 
(\ref{eq:WPDE}) that can be reached through an $\omega$--ansatz (\ref{eq:ansatz})
 by some $\mtilde$, $\ltilde$, $R$ satisfying, say, equation 
(\ref{eq:a}), there always exist three corresponding functions
$\mtilde$, $\ltilde$ and $R$ that satisfy the other three equations
(\ref{eq:b}) to (\ref{eq:d}) respectively, and lead to the same
solution $W_{\omega}$ of the differential equation 
(\ref{eq:WPDE}). 
For a proof the reader is referred to Appendix A.

Having shown the equivalence of the four ``linearized components'' of
the full non-linear equation ({\ref{eq:square}), we concentrate on the
most symmetric of them, equation (\ref{eq:a}), and think of it as
the ``representative'' of the whole set of equivalent equations
(\ref{eq:a})--(\ref{eq:d}). The reason we single out (\ref{eq:a}) is
its direct relation to the  action principle of the next section;
 it is therefore  important for the proper
foundation of any further developments to know the exact range of
equation (\ref{eq:a}), or, in other words, to compare its solutions
with the general solution of each $\omega$--equation
(\ref{eq:WPDE}). The rather surprising result of such a comparison is
that---for all weights---equations (\ref{eq:WPDE})  and (\ref{eq:a}) are exactly
equivalent; the remaining of this section is devoted to the relevant proof.

\subsection{Equivalence between the ``representative'' and each
weight--$\omega$ equation.}
The ansatz relation (\ref{eq:ansatz}) can be considered 
as a parametrised map, $a_{\omega}$, from  the set of functions $(\mtilde,\ltilde, R)$
satisfying equation (\ref{eq:a}) to the set of all solutions $W_{\omega}$ of
the corresponding 
weight--$\omega$ differential equation, with $W_{\omega}$, $W_{{\omega}H}$ and
$W_{{\omega}F}$ being different from zero.
The reason we have
included $R$ in the set of functions $(\mu,\lambda,R)$ is that
although $R$ is a function of $\mu$---defined by equation
(\ref{eq:root})---is not fully
specified by $\mu$ due to the sign ambiguity.

One would like to know whether the map $a_{\omega}$ is
one-to-one and, most importantly, whether it is onto. To check the
latter,  one
supposes that ${\mit v}$ is any solution of the $\omega$--equation
(\ref{eq:WPDE}), where---for simplicity---the subscript 
$\omega$ of ${\mit v}$ is omitted. A set of functions  $(\mtilde$, $\ltilde$, $R)$ obeying the ``representative'' equation
(\ref{eq:a}) is therefore required, with the property of 
producing through the $a_{\omega}$ map the
given solution ${\mit v}$. This is similar to the procedure followed in Appendix A in order to show that equations (\ref{eq:a})--(\ref{eq:d}) are equivalent; 
the difference is that the requirement that 
 at least one of the cases (\ref{eq:a})--(\ref{eq:d}) leads to 
${\mit v}$ is lifted---${\mit v}$  is now only confined to obey
equation (\ref{eq:WPDE}) for some weight $\omega$.

Clearly, the three conditions that  must be  satisfied  are:
\begin{enumerate}
\item The original differential equation
\begin{equation}
{\omega\over2}{\mit v}{\mit v}_F=\ftilde{\mit v}_F^2-{1\over 4}
{\mit v}_H^2,\qquad {\mit v}\neq 0\quad 
{\mit v}_F\neq 0\quad {\mit v}_H\neq 0.
\end{equation}
\item  The ``representative'' equation
\begin{equation}
{1\over\ltilde}\lambda_F
-{1\over\rtilde}\mu_F=0
\quad{\rm and}\quad
{1\over\ltilde}\lambda_H
-{1\over\rtilde}\mu_H=0;
\qquad
R=\sqrt{\bigl(\htilde-\mu\bigr)^2-\ftilde}.
\end{equation}
\item The ansatz relation
\begin{equation}
{\it v}=\ltilde^{\omega\over2}\biggl(\htilde-\mtilde+ \sqrt{(\htilde-\mtilde)^2-\ftilde}
\biggr)^{\omega\over2}.\label{eq:36}
\end{equation}
\end{enumerate}
The third condition can be written as 
\begin{equation}
{\it v}^{2\over{\omega}}=\ltilde\biggl(\htilde-\mtilde+ \sqrt{(\htilde-\mtilde)^2-\ftilde}
\biggr),
\label{eq:366}
\end{equation}
where this relation is valid up to a $2\over{\omega}$ power of
unity.   
Equation (\ref{eq:366}) can now be solved for $\mtilde$,  resulting in
\begin{equation}
\mtilde=\htilde-{1\over2}\Bigl({{\mit v}^{2\over{\omega}}\over\ltilde}+
{\ltilde\over {\mit v}^{2\over{\omega}}}\ftilde\Bigr).
\label{eq:3a}
\end{equation}
Differentiating $\mtilde$  with respect to both $\htilde$ and
$\ftilde$ gives
\begin{eqnarray}
\mtilde_H&=&1-{1\over\omega}
\Biggl      ({{\mit v}^{{2-\omega}\over\omega}\over\ltilde}-{\ltilde\over {\it v}^{{2+\omega}\over\omega}}\ftilde\Biggr) 
{\it v}_H+{1\over2}
\Biggl({{\it v}^{2\over\omega}\over\ltilde^2}-{1\over {\it v}^{2\over\omega}}\ftilde\Biggr)
\ltilde_H \qquad{\rm and}\nonumber\\  
\mtilde_F&=&-{1\over2}{\ltilde\over {\it v}^{2\over\omega}}
-{1\over\omega}
\Biggl({{\mit v}^{{2-\omega}\over\omega}\over\ltilde}-{\ltilde\over {\it v}^{{2+\omega}\over\omega}}\ftilde\Biggr)
{\it v}_F+{1\over2}
\Biggl({{\it v}^{2\over\omega}\over\ltilde^2}-{1\over {\it v}^{2\over\omega}}\ftilde\Biggr)
\ltilde_F.\label{eq:3c}
\end{eqnarray}
Conditions 2 and 3---being now in the same form---can easily be compared; more precisely, substituting equation
(\ref{eq:3a}) into  the expression for $R$---used in condition 2---one finds that 
\begin{equation}
\rtilde={1\over2}
\Biggl({{\it v}^{2\over\omega}\over\ltilde}-{\ltilde\over {\it v}^{2\over\omega}}\ftilde\Biggr),
\end{equation}
and hence  condition  2 becomes
\begin{equation}
\mu_H={1\over2}
\Biggl({{\it v}^{2\over\omega}\over\ltilde^2}-{1\over {\it v}^{2\over\omega}}\ftilde\Biggr)
\lambda_H
\qquad{\rm and}\qquad
\mu_F={1\over2}
\Biggl({{\it v}^{2\over\omega}\over\ltilde^2}-{1\over {\it v}^{2\over\omega}}\ftilde\Biggr)
\lambda_F.
\label{eq:2c}
\end{equation}

When equations (\ref{eq:2c})---derived from the second condition---are compared to 
equations (\ref{eq:3c})---derived from the third
condition---they lead to the following pair of equations for $\ltilde$:
\begin{equation}
\omega-\Biggl({{\it v}^{{2-\omega}\over\omega}\over\ltilde}-{\ltilde\over {\it v}^{{2+\omega}\over\omega}}\ftilde\Biggr)
{\it v}_H=0
\qquad{\rm and}\qquad
{\ltilde\over {\it v}^{2\over\omega}}+{2\over\omega}
\Biggl({{\it v}^{{2-\omega}\over\omega}\over\ltilde}-{\ltilde\over {\it v}^{{2+\omega}\over\omega}}\ftilde\Biggr)
{\it v}_F=0.
\label{eq:B}
\end{equation}
The above set of equations admits a common solution for $\ltilde$,
 we call it ${\bar\lambda}$,  given by 
\begin{equation}
{\bar\lambda}=-{2 {\it v}^{2\over\omega} {\it v}_F\over {\it v}_H}.
\label{eq:C}
\end{equation}
Equations (\ref{eq:B}) and (\ref{eq:C}) are all well defined since
${\it v}$, ${\it v}_H$, and ${\it v}_F$, are not allowed to vanish
identically, but for
the two equations in (\ref{eq:B}) to be consistent, they must lead either
to an identity or at least to a valid equation when
${\bar\lambda}$ is substituted into them---indeed, by doing so, they both reduce to
\begin{equation}
{\omega\over2}{\it v}{\it v}_F=\ftilde{\it v}_F^2-{1\over 4}{\it v}_H^2,
\end{equation}
which is of course true by virtue of condition 1. This proves that the
$a_{\omega}$ map
(\ref{eq:ansatz}) (from the set of functions $(\mtilde,\ltilde, R)$ to
the set of solutions of the corresponding $\omega$--equation
(\ref{eq:WPDE})) is {\it onto}.

The expression for  ${\bar\lambda}$ is now substituted back into equation 
(\ref{eq:3a}) to give the relevant expression  for $\mtilde$, 
\begin{equation}
{\bar {\mu}}=\htilde+{{\mit v}_F\over 
{\mit v}_H}\ftilde+{1\over4}{{\mit v}_H
\over {\mit v}_F},
\label{eq:D}
\end{equation}
and the one for $\rtilde$,
\begin{equation}
{\bar R}= {{\mit v}_F\over {\mit v}_H} \ftilde
-{1\over 4}{{\mit v}_H\over{\mit v}_F}
\label{eq:E}.
\end{equation}
Note that the expression for ${\bar R}$ is sign--unambiguous.

To check if the map $a_{\omega}$ is one--to--one, two sets of 
functions $(\mtilde_1,\ltilde_1,R_1)$ and $(\mtilde_2,\ltilde_2,R_2)$
are considered. They are required 
 to satisfy the ``representative'' pair of equations (\ref{eq:a}) and
provide---through the $a_{\omega}$ map---the same solution ${\it v}$ of the original
$\omega$--equation. 
The problem is readily solved  in Appendix A and leads to the
following three conditions,

\begin{eqnarray}
\ltilde_1^{\omega\over2}(\htilde-\mtilde_1+\rtilde_1)^{\omega\over2}&=&\ltilde_2^{\omega\over2}
(\htilde-\mtilde_2+\rtilde_2)^{\omega\over2},
\nonumber\\
{\ltilde_1^{\omega\over2}\over\rtilde_1}(\htilde-\mtilde_1+\rtilde_1)^{\omega\over2}&=&
{\ltilde_2^{\omega\over2}\over\rtilde_2}
(\htilde-\mtilde_2+\rtilde_2)^{\omega\over2},
\nonumber\\
{\ltilde_1^{\omega\over2}\over\rtilde_1}(\htilde-\mtilde_1+\rtilde_1)^{{\omega-2}\over2}      &=&
{\ltilde_2^{\omega\over2}\over\rtilde_2}(\htilde-\mtilde_2+\rtilde_2)^{{\omega-2}\over2},
\end{eqnarray}
which admit the almost trivial solution
\begin{equation}
\mtilde_1=\mtilde_2\qquad\ltilde_1^{\omega\over2}=\ltilde_2^{\omega\over2}\qquad R_1=R_2.
\end{equation}

The word ``almost'' is used because of the ambiguity in the expression
for the  $\lambda$'s. However, if the equivalence class of $\lambda$ is
defined as the set of functions that differ
from $\lambda$ by an $\omega\over2$ power of unity (it can be easily shown
that this defines an equivalence relation) then each anzatz--map (\ref{eq:ansatz})
becomes {\it one--to--one}.
Hence $\bar\mu$, $\bar R$ and the equivalence
class of $(\bar\lambda)$---given respectively by equations
(\ref{eq:D}), (\ref{eq:E}) and (\ref{eq:C})---are {\it unique}.

We have thus shown that the general solution of the
representative equation (\ref{eq:a}) is indeed equivalent to  the
general solution of the original arbitrary--weight differential equation
(\ref{eq:WPDE}) (with $W_{\omega}$,
$W_{{\omega}H}$, $W_{{\omega}F}$ not identically zero), and this completes the proof.

\subsection{On the local equivalence between the constraints.}
As a final remark, we return to one to the introductory comments of this
section, concerning the local equivalence of the old and new constraints---$(\ham,\mom)$ and $({\cal W}_{\omega},\mom)$ respectively---in case
that the combinations $W_{\omega}[\htilde,\ftilde]$ are used in vacuum
gravity. Specifically, for the constraint surface to be locally mapped  
onto itself under the replacement of $\ham$ by ${\cal W}_{\omega}[\htilde,\ftilde]$, the statement $W_{\omega}[\htilde,\ftilde]=0$ must be equivalent to the condition $\htilde=0$ when $\ftilde=0$.  
Looking back in equation (\ref{eq:ansatz}) however, it is fairly
transparent that at least for one choice of sign for the square root,
the constraint $W_{\omega}[\htilde,\ftilde]=0$ is satisfied {\it identically} at $\ftilde=0$, and hence does not enforce the necessary for equivalence Hamiltonian constraint. 

One would thus roughly expect---for each weight---half of the candidate
combinations $W_{\omega}[\htilde,\ftilde]$ to be inadequate for use in
vacuum gravity. Fortunately, the sign--unambiguous expression for
${\bar R}$, equation (\ref{eq:E}), makes certain that this is not
the case since---for $\ftilde=0$ and by virtue of equation (\ref{eq:D})---the quantities ${\bar R}$ and $\htilde$ - $\bar{ \mu}$ always come with the same sign and do not identically vanish. 

In particular, for any given solution ${\mit v}[\htilde,\ftilde]$ of the
weight--$\omega$ differential equation, the corresponding expression $\htilde
- \bar{ \mu} + {\bar R}$ reduces to -${\mit v}_H / 2{{\mit v}_F}$ at
$\ftilde=0$ and, therefore, there is no other
reason for ${\mit v}[\htilde,\ftilde]$ to identically vanish at
$\ftilde=0$, provided that ${\mit v}_H$[$\htilde$,0],
${\mit v}_F$[$\htilde$,0] and $(\bar{\lambda})$[$\htilde$,0]
are not identically zero themselves. 
It is of course still possible that ${\mit v}[\htilde,0]=0$ will not
imply the Hamiltonian constraint $\htilde=0$, but this cannot be known
unless the explicit form of a solution ${\mit v}[\htilde,\ftilde]$ is
given. An example of the above considerations concerns the 
Kucha\v{r}--Romano combination and is presented towards the end of the next section.

\section{An action functional for the weight--$\omega$ differential equation. }

The ``representative'' 
pair of equations (\ref{eq:a}) (being equivalent to the original
differential equation 
(\ref{eq:WPDE}) under the restriction that $W_{\omega}$,
$W_{{\omega}H}$ and $W_{{\omega}F}$ do not vanish identically) is of
particular interest because---as it will be shown in this section---can be naturally derived from an action principle involving a single 
scalar field and a single Lagrange multiplier with a non--derivative
coupling to gravity. 

For the physical relevance of the ``gravity plus scalar--field''
system, however, certain  reality conditions have to be taken into account.
They are imposed on almost every quantity used here and---because of
the ultimately close relation between the previous and the present section---they demand for an appropriate  modification of the results
of the first; this is done in section 4. 
Since  no other reference to some underlying
physical interpretation is present, the following  construction
should be viewed mainly as a mathematical one. 

\subsection{The scalar--field action.}
The relevant action functional $S^{\phi}$ is introduced as:
\begin{equation}
S^{\phi}[\phi, M, \gabu]=\int d^4X|\gamma|^{1\over2}\Biggl(
{1\over2}\lambda(M)\gabu\phi_{,\alpha}\phi_{,\beta}+\mu(M)\Biggr),
\label{eq:41}
\end{equation}
where the dependence of $\phi$, $M$ and $\gabu$ on the spacetime
points $X$ is not explicitly denoted.
The signature of the spacetime metric is taken to be $(-,+,+,+)$ and  $\lambda(M)$ and 
$\mu(M)$ are some given but otherwise arbitrary
continuous functions of the Lagrange multiplier
$M$. Extension to an arbitrary number of scalar fields and multipliers is possible, but
seems rather redundant. 

To ensure that the scalar field
$\phi$ will always be present in the action functional,
 $\lambda(M)$ is required to be different from zero---no such
restriction is imposed on $\mu(M)$. 
The notation for the two functions of the multiplier is
 intentionally chosen to reflect the notation used in section 2;  unlike
the corresponding
quantities there, however, $\lambda(M)$ and 
$\mu(M)$ must now be {\it real--valued}\footnote{Recall that this does not mean that $\lambda(M)$ and 
$\mu(M)$ must necessarily be {\it real} functions of $M$.} and they
must also come with the {\it correct sign}---this is further discussed
in subsection 3.2.

If the ``dimension'' of $\gabd dX^{\alpha}dX^{\beta}$ is defined as
length squared,
\begin{equation}
\bigl[\gabd dX^{\alpha}dX^{\beta}\bigr]=L^2,
\end{equation}
and the action $S^{\phi}$ is required to be dimensionless, 
the only consistent attribution of dimensions to the various terms
appearing in (\ref{eq:41})---keeping the conventional dimensions of
inverse length for the scalar field---is the following:
\begin{equation}
[\phi]=L^{-1},\qquad [M]=[\lambda(M)]=L^0=1,\qquad [\mu(M)]
=L^{-4}. 
\end{equation}
This means that $\mu(M(X))$ can be considered as a function of the
multiplier $M(X)$ scaled by a constant scalar function $C(X)$, 
\begin{equation}
\mu(M(X))=C(X)\rho(M(X)),
\end{equation}
where $[C]=L^{-4}$ and $[\rho(M)]=L^0=1$. For simplicity, appropriate
units can be chosen so that the value of
$C(X)$ equals 1. 

\subsection{A.D.M. decomposition of the coupled system.}
By  coupling  the field action $S^{\phi}[\phi, M, \gabu]$ to 
the gravitational Einstein-Hilbert action $S[\gabd]$
\begin{equation}
S[\gabd]=\int d^4X|\gamma|^{1\over2} R[\gabd],
\end{equation}
and by proceeding with the usual A.D.M. decomposition \cite{ADM} of
the total action,
\begin{equation}
S^T:=S+S^{\phi},
\end{equation}
one obtains the constraints $\hamtot$ and $\momtot$, whose 
 form is common to any theory with a non-derivative coupling to
gravity \cite{HoKuTe}: 
\begin{eqnarray}
\hamtot &:=&\ham +\hamphi =0
\label{eq:hamt}\\
\momtot &:=&\mom+\momphi=0.
\label{eq:momt}
\end{eqnarray}
The gravitational parts of the constraints, $\ham$ and $\mom$, are identical 
to the constraints of vacuum general relativity (written out in
equations (\ref{eq:constraint})), while the field contributions $\hamphi$ and $\momphi$ are like the ones of a massless
scalar field, only extended by the presence of the two functions
$\lambda(M)$ and $\mu(M)$,
\begin{eqnarray}
\momphi &=&\pi\phi_{,i}
\label{eq:mphi}\\
\hamphi &=&g^{1\over2}\Biggl(-{1\over2}{\pi^2\over g\lambda(M)}-\mu(M)
-{1\over2}{\lambda(M)\over\pi^2}g^{ij}\momphi\momjphi\Biggr),
\label{eq:hphi}
\end{eqnarray}
where $\pi$ is the momentum conjugate to the field $\phi$. 

From equation
(\ref{eq:hphi}) one observes that $\lambda(M)$ must be restricted to be {\it
negative--valued}, in order for the scalar field to have positive kinetic
energy. On the other hand, and in search of a proper interpretation for
$\mu(M)$, we make use of the fact that $\mu(M)$ appears as a cosmological
constant in equation (\ref{eq:hphi}), and
allow it
to have any {\it real} value at
all. It must be said, however, that this uncertainty about the correct sign
of $\mu(M)$ is rather unimportant (if such thing as a ``correct'' sign
really exists), since care is also taken to make the
remaining of this discussion essentially independent of the actual
restrictions imposed. The ``negativity'' of $\lambda(M)$ and
the ``reality'' of $\mu(M)$ constitute the {\it first reality condition} presented in this section. 

\subsection{The two equations for $M$ and $\pi$.}
At this stage, the total action $S^T$ can be varied with respect to the
multiplier. This only appears in the field action $S^{\phi}$ and
in particular in the $\hamphi$--term, when the latter is written in A.D.M. form:
\begin{equation}
S^{\phi}=\int d^3xdt\biggl(\pi\dot\phi-N\hamphi-N^i\momphi\biggr).
\end{equation}
As a result, $M$ can be equivalently determined by requiring that
\begin{equation}
{d\hamphi\over d M}=0,
\end{equation}
which produces the following condition:
\begin{equation}
{1\over2}{\pi^2\over g\lambda^2(M)}\lambda'(M)
-\mu'(M)-{1\over 2}{1\over\pi^2}\lambda'(M)g^{ij}\momphi\momjphi=0.
\label{eq:zero}
\end{equation}
In a usual notation, $\lambda'(M)$ and $\mu'(M)$ denote the total derivatives of $\lambda(M)$ and $\mu(M)$ with respect to $M$. 
The constraints (\ref{eq:hamt}) and (\ref{eq:momt}) can now be used
to rewrite equations (\ref{eq:hphi}), (\ref{eq:zero}) in terms of
the gravitational contributions to these constraints,
\begin{eqnarray}
{1\over2}{\pi^2\over g\lambda(M)}+{1\over2}{\lambda(M)\over\pi^2}
g^{ij}\mom\momj &=& \htilde-\mu(M)
\label{eq:13}\\
{1\over2}{\pi^2\over g\lambda^2(M)}\lambda'(M)
-{1\over2}{1\over\pi^2}g^{ij}\mom\momj\lambda'(M)&=& \mu'(M),
\label{eq:14}
\end{eqnarray}
where $\htilde$ is the zero--weight scalar density defined in \cite{Mark}.

The pair of equations (\ref{eq:13}) and (\ref{eq:14})---relating the quantities
$\pi$, $M$, $\ham$, and $\mom$---is the starting point of the main
discussion of this section. The aim is to solve these
equations for $\pi$ and $M$ in terms of $\ham$ and $\mom$, regarding the functions
$\lambda(M)$ and $\mu(M)$ as known. Because the analysis depends on the
actual form of the derivatives, one should distinguish some special cases---in which either $\lambda'(M)$ or $\mu'(M)$ is identically zero or both 
$\lambda'(M)$ or $\mu'(M)$ are identically zero---and
treat them separately from the general case where the
derivatives do not vanish.

\subsection{Solving for $M$ and $\pi$.}
\subsubsection{The general case. Recovery of the $\omega$--ansatz and
reconstruction of the ``representative'' equation.}
The general case then occurs when:
\begin{equation}
\lambda'(M)\neq 0\qquad\mu'(M)\neq 0.
\label{eq:biga}
\end{equation}
When this condition holds, one can  multiply equation (\ref{eq:14}) by $(\lambda(M)/\lambda'(M))$
and obtain its equivalent version
\begin{equation}
{1\over2}{\pi^2\over g\lambda(M)}-{1\over2}
{\lambda(M)\over\pi^2}g^{ij}\mom\momj=
{\mu'(M)\lambda(M)\over\lambda'(M)}.
\label{eq:15}
\end{equation}

As already stated, the aim is to solve equations (\ref{eq:13}) and 
(\ref{eq:15}) for $\pi$ and $M$ in terms of the gravitational contributions to the constraints. This can be done by adding and subtracting
(\ref{eq:13}) and (\ref{eq:15}), and then cross--multiplying the
resulting equations to eliminate the field momenta. One is left
with an algebraic equation determining the
multiplier $M$ in terms of $\ham$, $\mom$ alone,
\begin{equation}
{\mu'(M)\lambda(M)\over\lambda'(M)} =
\sqrt{\biggl(\htilde-\mu(M)\biggr)^2-\ftilde},
\label{eq:16}
\end{equation}
where both choices of sign for the square root are allowed, and $\ftilde$ is the quantity defined in \cite{Mark}. 
To get the corresponding expression for
the field momenta $\pi $ in terms of $\ham$ and $\mom$,
equation (\ref{eq:13}) is solved to give an expression 
of $\pi$ as a function of $M$, $\ham$, and
$\mom$, 
and then a solution of (\ref{eq:16}) is substituted in this 
expression, to replace $M$. This is subject to the problem of
existence of a proper (real or complex) solution for $M$ so that the
functions $\lambda(M)$ and $\mu(M)$ are negative--valued and
real--valued respectively; this problem, however, is finally
avoided in section 4. 

In particular, if $M[\ham,\mom]$ denotes such a proper solution, the
corresponding expression for the square of the field momentum---written in 
weightless form---is given by
\begin{equation}
{1\over(g^{{1\over2}})^2}\pi^2[\ham\mom]=\lambda(M[\ham,\mom])
\Biggl(\htilde-\mu(M[\ham,\mom])+
\sqrt{\Bigl(\htilde-\mu(M[\ham,\mom])\Bigr)^2-\ftilde}
\Biggr),
\label{eq:17}
\end{equation}
which, together with (\ref{eq:16}), provide the required set of solutions of
the original system of equations (\ref{eq:13}) and (\ref{eq:14}). 

One then observes that the actual form of equations (\ref{eq:16}) and (\ref{eq:17})
 ensures  that the Hamiltonian and momentum contributions to the constraints, $\ham$ and $\mom$, 
only appear in the form of the scalar combinations $\htilde$ and $\ftilde$
respectively. This means that the solutions $M[\ham,\mom]$ and
$\pi^2[\ham,\mom]$ can be written as $M[\htilde,\ftilde]$ and
$\pi^2[\htilde,\ftilde]$, and also by regarding $\lambda$ and $\mu$ as
functions of $\htilde$ and $\ftilde$, according to 
\begin{equation}
\ltilde[\htilde,\ftilde]:=\lambda(M[\ham,\mom])
\qquad{\rm and}\qquad
\mtilde[\htilde,\ftilde]:=\mu(M[\ham,\mom]),
\label{eq:18}
\end{equation}
we can bring equation  (\ref{eq:17}) into the equivalent form 

\begin{equation}
{1\over(g^{{1\over2}})^2}\pi^2[\htilde,\ftilde]=
\ltilde[\htilde,\ftilde]
\Biggl(\htilde-\mtilde[\htilde,\ftilde]+
\sqrt{\Bigl(\htilde-\mtilde[\htilde,\ftilde]\Bigr)^2-\ftilde}
\Biggr).
\label{eq:19}
\end{equation}

If the above expression for the field momentum is raised to the power of $\omega\over2$, is immediately
recognised as the ansatz equation (\ref{eq:ansatz}) of section 2.  It satisfies the differential  equation
(\ref{eq:WPDE}), provided that $\ltilde[\htilde,\ftilde]$,
$\mtilde[\htilde,\ftilde]$ and the square root in equation
(\ref{eq:19}) obey the---common to all weights---``representative'' equation (\ref{eq:a}). Quite remarkably,
the last statement is true as the following argument demonstrates:
 
Since (\ref{eq:16}) is an algebraic equation for $M$, it
must hold identically when written in terms of an actual (real or
complex) solution 
$M[\htilde,\ftilde]$. As a result, it automatically becomes a
{\it differential} equation for $\lambda(M[\htilde,\ftilde])\equiv\ltilde[\htilde,\ftilde]$
and $\mu(M[\htilde,\ftilde])\equiv\mtilde[\htilde,\ftilde]$, whatever
the functions
$\lambda(M)$ and $\mu(M)$ were initially chosen to be. Furthermore, equation (\ref{eq:16}) makes certain that its solution
$M[\htilde,\ftilde]$ will always satisfy\footnote{Again, this means
that $M_H$ and $M_F$ must not identically vanish, but proper care
should be taken to restrict to regions of the $\htilde$, $\ftilde$
plane where $M_H$ and $M_F$ do not even take the {\it value} zero.}
\begin{equation}
M_H\neq 0
\qquad{\rm and}\qquad
M_F\neq 0
\label{eq:20}
\end{equation}
and, therefore, by multiplying equation (\ref{eq:16}) with 
$M_H$ and $M_F$, we get a pair of two partial differential equations,
\begin{equation}
{1\over\ltilde}\lambda_H-
{1\over\sqrt{(\htilde-\mtilde)^2-\ftilde}}
\mu_H=0
\qquad{\rm and}\qquad
{1\over\ltilde}\lambda_F-
{1\over\sqrt{(\htilde-\mtilde)^2-\ftilde}}
\mu_F=0,
\label{eq:21}
\end{equation}
which is exactly the ``representative'' equation (\ref{eq:a}). 

In other words, for any initial choice (\ref{eq:biga}) of 
functions $\lambda(M)$ and $\mu(M)$ (subject to the condition that there exist a proper solution
for $M$ that makes $\lambda(M)$ and $\mu(M)$ negative--valued and real--valued) the
action (\ref{eq:41}) necessarily produces    
arbitrary--weight combinations
of the gravitational constraints that have the property of
generating a true Lie algebra; they are explicitly given by the $\omega\over2$
power of the square of the field momentum, equation (\ref{eq:19}). 

Note, however,
that for equation (\ref{eq:19}) to make sense, there must exist some regions of
the $\htilde$, $\ftilde$ plane---equivalently, regions of the
gravitational phase space---such that the quantity inside the square root, as well as the
whole right hand side of equation (\ref{eq:19}) are {\it
positive--valued}. This is the  
{\it second reality--condition} of the section which---as the first
one---is taken into proper account in section 4.

We can now return to our starting point---equations
(\ref{eq:13}), (\ref{eq:14})---and treat the special cases 
where either $\lambda'(M)$ or $\mu'(M)$ is identically zero, or both
of them are identically 
zero. There are three possibilities:
\subsubsection{Special case 1. The ``Kucha\v{r}--Romano'' family.}

\begin{equation}
\lambda'(M)=0\qquad{\rm and}\qquad\mu'(M)=0.
\label{eq:lzmz}
\end{equation}
Suppose that $\mu(M)=C_1$ and $\lambda(M)=C_2$, where $C_1$ and
$C_2$ are, respectively, real and negative constants, according to the first
reality condition. Now there is no
multiplier present in the total action and, therefore, equation
(\ref{eq:14}) is trivially satisfied, both sides being equal to
zero. Correspondingly, the coupled system of equations (\ref{eq:13}), (\ref{eq:14}) for
$M$ and $\pi^2$ reduces to the single equation (\ref{eq:13}),
giving $\pi^2$ directly as a function of $\htilde$ and $\ftilde$:

\begin{equation}
{\pi^2\over (g^{1/2})^2}=C_2\Biggl(\bigl(\htilde-C_1\bigr)
+\sqrt{\bigl(\htilde-C_1\bigr)^2-\ftilde}\Biggr).
\label{eq:25}
\end{equation}
When this is raised to the $\omega\over2$ power, it has the form
of the $\omega$--ansatz equation (\ref{eq:ansatz}), provided that the identification 
\begin{eqnarray}
\ltilde[\htilde,\ftilde]&\equiv&C_2\\
\mtilde[\htilde,\ftilde]&\equiv&C_1
\end{eqnarray}
is made.
The ``representative'' equation, (\ref{eq:21}) or (\ref{eq:a}), is satisfied trivially for these
$\lambda[\htilde,\ftilde]$ and $\mu[\htilde,\ftilde]$,
and therefore expression (\ref{eq:25}) provides further solutions of the
differential equation (\ref{eq:WPDE}); they are also required to be
positive--valued, 
according to the second reality condition.  

This case is interesting because it reduces to the Kucha\v{r}--Romano 
combination under the identification $\omega=2$, $C_1=0$ and
$C_2=-1$. It also illustrates our previous comment---made at the end of
section 2---concerning the equivalence of the old and new constraints, in
case that the Hamiltonian constraint is replaced by any
self--commuting constraint combination. More precisely,
from the two different weight--two solutions defined by equation
(\ref{eq:25}) (corresponding to the two choices of sign for the square
root) one obeys
${\mit v}_H$ [$\htilde$,0]$=0$ and according to subsection 2.4 is excluded,
while the other one---although it satisfies ${\mit v}_H$
[$\htilde$,0]$\neq0$, ${\mit v}_F$ [$\htilde$,0]$\neq0$ and
$(\bar{\lambda})[\htilde,0] \neq 0$---does not imply the Hamiltonian
constraint when evaluated at $\ftilde=0$, unless the (real) constant
$C_1$ is chosen to be zero. This surviving combination is of course one of the two $\Lambda_{\pm}$.  

\subsubsection{Special case 2. The ``pseudomultiplier''.}

\begin{equation}
\lambda'(M)=0\qquad{\rm and}\qquad\mu'(M)\neq 0.
\label{eq:lzmnz}
\end{equation}
Suppose that $\lambda(M)=C_2$, where $C_2$ is negative. Equation (\ref{eq:14})
becomes\footnote{To avoid any possible confusion we point out that 
equations (\ref{eq:lzmnz}) and (\ref{eq:mz}) are not contradictory; 
the first means that $\mu'(M)$ must not be {\it identically} zero while the 
second is an {\it algebraic} equation for determining $M$ if one regards $
\mu(M)$ as known.}
\begin{equation}
\mu'(M)=0.
\label{eq:mz}
\end{equation}
Any possible solution of this equation can
only produce a {\it numerical} value for $M$ and, therefore, $M$ is not a proper
multiplier but merely ``fixes itself a value''. However, one still
proceeds by solving (\ref{eq:13}) for $\pi^2$ and finds
\begin{equation}
{\pi^2\over (g^{1/2})^2}=-C_2\Biggl(\bigl(\htilde-C_1\bigr)
+\sqrt{\bigl(\htilde-C_1\bigr)^2-\ftilde}\Biggr),
\end{equation} 
where now $C_1$ is the real numerical value of $\mu(M)$ after the
elimination of the ``pseudomultiplier'' $M$. Therefore, provided that
a proper solution for $M$ (leading to a real--valued $\mu(M)$) exists, case 2 is
essentially equivalent to case 1. 
\subsubsection{Special case 3. The ``null vector'' family.} 
 
\begin{equation}
\lambda'(M)\neq0\qquad{\rm and}\qquad\mu'(M)= 0.
\label{eq:lnzmz}
\end{equation}
Suppose that $\mu(M)=C_1$, $C_1$ being real. Now equation (\ref{eq:14}) becomes
\begin{equation}
\Biggl({\pi^2\over g\lambda(M)}-{\lambda(M)\over\pi^2}
g^{ij}\mom\momj\Biggr){\lambda'(M)\over\lambda(M)}=0,
\label{eq:30}
\end{equation}
and, as a result, $M$  either has a real numerical value (thus producing exactly the
same combinations as in special cases 1 and 2) or it satisfies
\begin{equation}
{\pi^2\over g\lambda(M)}-{\lambda(M)\over\pi^2}g^{ij}\mom\momj=0.
\label{eq:second}
\end{equation}
When (\ref{eq:second}) is combined with equation (\ref{eq:13}), leads to a 
constraint between the two variables $\ftilde$ and $\htilde$, namely
\begin{equation}
\ftilde=(\htilde-C_1)^2.
\label{eq:ctwo}
\end{equation}
An example is when
$\lambda(M)=M$ and $\mu(M)=0$. It can be interpreted as a coordinate
condition on $\gabd$ such that $\phi_{,\alpha}$ becomes a null vector,
\begin{equation}
\gabu\phi_{,\alpha}\phi_{,\beta}=0,
\end{equation}
provided  that $\phi$ is considered as an externally fixed field.

\section{The inverse procedure.}

It was demonstrated in the previous section that apart from some improper choices for $\mu(M)$ and $\lambda(M)$---resulting either in constraints  
(\ref{eq:ctwo}) between the two gravitational variables $\htilde$ and
$\ftilde$, or in solutions of equation
(\ref{eq:16}) that lead to complex--valued quantities---all other
choices yield, in principle, a solution
of the $\omega$--differential equation (\ref{eq:WPDE}). 
It is given by the $\omega\over2$ power of the square of the scalar field
momentum when the latter is evaluated in terms of the gravitational 
variables alone and---because of that---must necessarily admit real
positive values in some regions of the $\htilde$, $\ftilde$ plane.

The important question that we deal with in this section is
whether the totality of solutions $(\pi^{2})^{\omega\over
2}[\htilde, \ftilde]$ provides the {\it complete} ``real'' subset of solutions\footnote{Excluding the physically irrelevant cases when
$W_{\omega}[\htilde, \ftilde]=0$,
$W_{{\omega}H}[\htilde, \ftilde]=0$ and  $W_{{\omega}F}[\htilde,
\ftilde]=0$.} of the
weight--$\omega$ equation; the latter is meant to be the set of all solutions
of (\ref{eq:WPDE}) that satisfy the reality conditions of
section 3.

Using the ``onto'' property of each $\omega$--ansatz map (discussed in
section 2), the above question  is equivalent to asking if the
set of functions 
($\lambda(M[\htilde, \ftilde])$, $\mu(M[\htilde, \ftilde])$,
$R(M[\htilde, \ftilde])$) (obtained by the
action principle of the previous section) can provide the {\it
complete} ``real--sending'' subset of solutions of the ``representative'' 
equation; i.e., the (weight--dependent) set of solutions of the
``representative'' equation whose {\it image} under each 
$\omega$--ansatz map is the ``real''
subset of the corresponding $\omega$--equation. The re-phrasing is
essential because it allows  
a comparison between {\it linear} equations to take place, which
greatly 
simplifies the whole problem.

\subsection{The ``real'' subset of solutions.}

The exact definition of the ``real'' subset of solutions involves the following procedure:
For every solution $W_{\omega}[\htilde, \ftilde]$ of the
$\omega$--equation, one obtains the corresponding quantities
$\bar\lambda[\htilde, \ftilde]$ and $\bar\mu[\htilde, \ftilde]$, that
are given respectively by equations (\ref{eq:C}) and
(\ref{eq:D}). For our present purposes, two triplets of functions
($W_{\omega}$, $\bar\lambda$,
$\bar\mu$) and ($W_{\omega}$, $\bar\lambda'$,
$\bar\mu$) are considered to be {\it different} if $\bar\lambda$
and $\bar\lambda'$ differ by an $\omega\over2$ power
of unity, unless of
course this $\omega\over2$ power is unity itself.

The label ${\cal{Q}}_{W_{\omega}}$ is then attached to every possible triplet ($W_{\omega}$, $\bar\lambda$,
$\bar\mu$), and is defined as the {\it largest set} of values of
$\htilde$ and $\ftilde$ for which $W_{\omega}$ and
the corresponding quantities $\bar\lambda$
and $\bar\mu$ are {\it positive}, {\it negative} and
{\it real--valued},
respectively.  In other words, given any solution $W_{\omega}$, its
label ${\cal{Q}}_{W_{\omega}}$ provides the regions of the
gravitational phase space where all the reality conditions of section
3 are
satisfied. It should be mentioned, however, that although the
labeling is {\it suggested} by the results of section 3, it {\it does not} depend on it {\it in any
way}---if it did, the whole procedure would simply be inconsistent.   

The ``real'' subset of solutions of each $\omega$--equation can then be 
defined as the set of solutions $W_{\omega}$, for
which the label ${\cal{Q}}_{W_{\omega}}$ of the  
corresponding triplet ($W_{\omega}$, $\bar\lambda$,
$\bar\mu$) is {\it not the empty set}. The
definition is clearly independent of the actual reality conditions
imposed---in the sense that it can be trivially adapted to more refined
physical conditions---and thus justifies a previous comment in
subsection 3.2. In addition, because of
the requirement on $\bar\lambda$ to have a unique sign, only {\it one}
member (at most) of each former equivalence class 
can survive the definition of reality, which means that the  
``real--sending'' subset is {\it still} mapped onto the ``real'' subset in an {\it one--to--one}
fashion.

\subsection{The inverse procedure: Action functionals
for given ``real'' solutions.}

Using the above terminology, the main question of the section is
addressed once more, whether the functions ($\lambda(M[\htilde, \ftilde])$,
$\mu(M[\htilde, \ftilde])$, $R(M[\htilde, \ftilde])$)---obtained by the action prescription of
section 3---can generate the {\it complete} 
``real--sending'' subset of the ``representative'' equation.

In case they did generate the complete ``real--sending'' subset, all problems 
concerning the improper choices for  $\mu(M)$  and  $\lambda(M)$---mentioned in the begining of this section---could be easily considered as irrelevant, since now for 
every ``real'' solution $W_{\omega}[\htilde,\ftilde]$ of the weight--$\omega$
differential equation there would always exist a proper choice of
$\mu(M)$ and $\lambda(M)$ that would lead to this solution.  

In particular, for the above statement to be true, there must exist some (real or complex) functions $M[\htilde,\ftilde]$, $\mu(M)$ and 
$\lambda(M)$ that satisfy the conditions
\begin{eqnarray}
\mu\Bigl(M[\htilde,\ftilde]\Bigr)&=&\bar\mu[\htilde,\ftilde]\label{eq:x}\\
\lambda\Bigl(M[\htilde,\ftilde]\Bigr)&=&{\bar\lambda}[\htilde,\ftilde]\qquad{\rm
and}\label{eq:y}\\
\sqrt{\Bigl(\htilde-\mu\bigl(M[\htilde,\ftilde]\bigr)\Bigr)^2-
\ftilde}&=&\bar R[\htilde,\ftilde]\label{eq:z},
\end{eqnarray}
for every ``real'' solution $W_{\omega}[\htilde,\ftilde]$.  
The overbar symbol on the right hand side of
equations (\ref{eq:x}), (\ref{eq:y}), (\ref{eq:z}) is a reminder
of the uniqueness of these expressions for each given
$W_{\omega}[\htilde,\ftilde]$, as explained in
subsection 4.1.

By inspecting the representative equation (\ref{eq:a})---which holds for any such set of functions ($\bar\lambda$, $\bar\mu$, $\bar R$)---one observes that $\bar{\lambda}[\htilde,\ftilde]$ and $\bar{\mu}[\htilde,\ftilde]$ can either be 
functions of {\it both} $\htilde$ and $\ftilde$ or {\it constants}; if this is not so, the system of the two partial differential 
equations in (\ref{eq:a}) is self--contradictory. Correspondingly, we have to
distinguish between the two cases:

\subsubsection{The special case. Constant functions
$\bar{\lambda}[\htilde,\ftilde]$ and $\bar{\mu}[\htilde,\ftilde]$.}
If $\bar\lambda$[$\htilde$,$\ftilde$] and 
$\bar\mu$[$\htilde$,$\ftilde$] are negative and
real--valued constants, say $C_2$ and
$C_1$ respectively, then special case 1 in subsection 3.4.2 suggests 
that---without needing to specify
the Lagrange multiplier as a function $M$[$\htilde$,$\ftilde$]
of the gravitational variables---one can directly identify the
required functions
\begin{equation}
\lambda(M)=\bar\lambda[\htilde,\ftilde]=C_2
\qquad{\rm and}\qquad
\mu(M)=\bar\mu[\htilde,\ftilde]=C_1.
\end{equation}
The above relations satisfy equation (\ref{eq:z}) for an 
appropriate choice of sign and, therefore, the problem of 
finding an action functional is solved.

\subsubsection{The general case. Non--trivial functions
$\bar{\lambda}[\htilde,\ftilde]$ and $\bar{\mu}[\htilde,\ftilde]$ and
the unique ``$\kappa$''.}
In the general case, when $\bar\mu[\htilde,\ftilde]$ 
and $\bar\lambda[\htilde,\ftilde]$ are non--trivial 
functions of $\htilde$ and $\ftilde$, the situation is
more complicated and the following four conditions must be satisfied:
\begin{eqnarray}
\bar\mu_H[\htilde,\ftilde]=\mu'(M) M_H[\htilde,\ftilde]
&\qquad&
\bar\mu_F[\htilde,\ftilde]=\mu'(M) M_F[\htilde,\ftilde]\nonumber\\
\bar\lambda_H[\htilde,\ftilde]=\lambda'(M) M_H[\htilde,\ftilde]
&\qquad&
\bar\lambda_F[\htilde,\ftilde]=\lambda'(M) M_F[\htilde,\ftilde].
\label{eq:mukail}
\end{eqnarray}
From (\ref{eq:mukail}) one gets a differential equation 
for the required $M[\htilde,\ftilde]$,
\begin{equation}
M_H[\htilde,\ftilde]-{\cal A}[\htilde,\ftilde] M_F[\htilde,\ftilde]=0,
\label{eq:colo}
\end{equation}
where 
\begin{equation}
{\cal A}[\htilde,\ftilde]={\bar\mu_H\over\bar\mu_F}=
{\bar\lambda_H\over\bar\lambda_F}.
\label{eq:cala}
\end{equation}
Note that because 
of the properties of the ``representative'' equation, the denominator and
the numerator in the above expression do not identically vanish and 
therefore---provided that all ``improper'' 
regions of the ($\htilde, \ftilde$) plane are excluded---equation (\ref{eq:cala}) is in general well--defined.

An explicit expression for ${\cal A}[\htilde,\ftilde]$ (in 
terms of a given weight--$\omega$ solution ${\mit v}[\htilde,\ftilde]$ and 
its derivatives) is found by substituting equations (\ref{eq:D}, \ref{eq:C}) for 
$\bar\mu$ and $\bar\lambda$ into equation (\ref{eq:cala}): 
\begin{equation}
{\cal A}[\htilde,\ftilde]={{\mit v}_H^2{\mit v}_F+{\mit v}{\mit v}_H{\mit v}_{HF}-
{\mit v}{\mit v}_F{\mit v}_{HH}\over
{\mit v}_F^2{\mit v}_H+{\mit v}{\mit v}_H{\mit v}_{FF}-
{\mit v}{\mit v}_F{\mit v}_{HF}},
\end{equation}
where ${\mit v}_{HH},\ {\mit v}_{HF}$ and $ {\mit v}_{FF}$ denote the
second partial derivatives of  ${\mit v}$ with respect to $\htilde$
and $\ftilde$. Note that (4.8) is $\omega$--independent.

By examining equation (\ref{eq:colo}), 
it can be seen that any arbitrary (real or complex) function $f$ of
$\bar\mu,\bar\lambda$ provides a solution to it: 
\begin{equation}
M[\htilde,\ftilde]=f\bigl(\bar\mu[\htilde,\ftilde], \bar\lambda[\htilde,\ftilde]\bigr).
\label{eq:S}
\end{equation}
An important observation, however, is that---due to the ``representative''
equation (\ref{eq:a})---the Jacobian of $\bar\mu[\htilde,\ftilde]$ and
$\bar\lambda[\htilde,\ftilde]$ with respect to the variables 
$\htilde,\ftilde$ is identically zero; hence there are 
at least some local regions of the $\htilde$, $\ftilde$ 
plane where $\bar\mu[\htilde,\ftilde]$ is solvable as some 
 {\it unique real} ``$\kappa$'' function of $\bar\lambda[\htilde,\ftilde]$: 
\begin{equation}
\bar{\mu}[\htilde,\ftilde]=\kappa\bigl(\bar{\lambda}[\htilde,\ftilde]\bigr).
\label{eq:skoto}
\end{equation} 
Since $\bar\mu$ and $\bar\lambda$ only depend on the specific
solution $W_{\omega}[\htilde,\ftilde]$, the same is true for the uniquely defined
$\kappa$. 
Equation (\ref{eq:S}) then reduces to
\begin{equation}
M[\htilde,\ftilde]=h(\bar\lambda[\htilde,\ftilde]),
\label{eq:pret}
\end{equation}
where $h$ is some arbitrary (real or complex) function which, in addition, can be
chosen invertible.
 
Having obtained an expression (\ref{eq:pret}) for the multiplier
$M[\htilde,\ftilde]$, one only needs to find the remaining required functions 
$\mu(M)$, $\lambda(M)$, with the property that 
\begin{eqnarray}
\mu\Bigl(h(\bar\lambda[\htilde,\ftilde)\Bigr)&=&\bar\mu[\htilde,\ftilde]
=\kappa(\bar\lambda[\htilde,\ftilde])\nonumber\\
\lambda\Bigl(h(\bar\lambda[\htilde,\ftilde)\Bigr)&=&\bar\lambda[\htilde,\ftilde].
\label{eq:muandl}
\end{eqnarray}
From (\ref{eq:muandl}) these functions are easily found to be
\begin{eqnarray}
\mu(M)=\kappa\circ h^{-1}(M)\nonumber\\
\lambda(M)=h^{-1}(M),
\label{eq:www}
\end{eqnarray}
which are well--defined since $h$ can be always chosen 
invertible. 

An equivalent way of writing (\ref {eq:www})---which can be seen directly from equation (\ref{eq:skoto})---is   
\begin{equation}
\mu(M)=\kappa\bigl(\lambda(M)\bigr),
\label{eq:ppp}
\end{equation}
where $\lambda$ is kept arbitrary. 
This expression---which now only involves the $\kappa$ function---is the required solution
to our problem and {\it justifies the title of the section}. 
It is reminded that $\kappa$ is specified uniquely by the given solution $W_{\omega}[\htilde,\ftilde]$ of the
weight--$\omega$ differential equation (at least locally).

\subsubsection{An application. The Brown and Kucha\v{r} combination:} 

As an example, the Brown and Kucha\v{r} 
combination  $W_2[\htilde,\ftilde]=\htilde^2-\ftilde$ (in its
weightless form) is considered.
By using equations (\ref{eq:D}) and (\ref{eq:C}), one finds 
the corresponding unique expressions $\bar\mu$ and $\bar\lambda$ as 
\begin{equation}
\bar\mu={1\over2}\Biggl(\htilde-{\ftilde\over\htilde}\Biggr)\
\qquad{\rm and}\qquad
\bar\lambda=\Biggl(\htilde-{\ftilde\over\htilde}\Biggr).
\label{eq:halfs}
\end{equation}
The set of values of $\htilde$ and $\ftilde$ for which the two reality
conditions are satisfied (i.e., the label ${\cal{Q}}_{W_2}$) is given 
by the two inequalities $\htilde^2 > \ftilde$ and $\htilde < 0$ and,
therefore, the
combination $W_2$ indeed belongs to the
``real'' subset of solutions of the weight--two equation. 
Furthermore, for this range of values, the quantities $W_2$, $\bar\lambda$,
$\bar\mu$, $\bar R$, as well as their derivatives are all
{\it well defined}, and thus the same also applies to the whole procedure in
sections 2 and 3. 

The uniquely specified real function $\kappa$ is then determined as
\begin{equation}
\kappa(\bar\lambda[\htilde,\ftilde])={1\over2}\bar\lambda[\htilde,\ftilde]
=\bar\mu[\htilde,\ftilde]
\end{equation}
and, therefore, any choice of
$\mu(M)=\kappa\bigl(\lambda(M))=\lambda(M)/2$ is fine, with  
$\lambda(M)$ being an arbitrary (real or complex) function of $M$. This illustrates a statement in the introduction of
the paper, that there is actually a {\it variety} of single--scalar--field
actions which lead to the combination $G(x)$\footnote{It seems
interesting to mention here that $G(x)$ is a very special kind of combination, since
the family it generates---by the algorithm $G_{\omega}(x)$ := $G^{{\omega}\over2}(x)$)---provides the {\it only} solutions of equation (\ref{eq:WPDE}) that are {\it
polynomial} in $\htilde$ and $\ftilde$.}.

\section{Future possibilities.}

In this paper, we attempted to find a phenomenological medium that
would produce constraint--combinations similar to the ones obtained by
algebraic means in \cite{Mark}, hoping that such a procedure would
throw more light on their origin. For this reason, we transformed the
nonlinear arbitrary--weight differential equation of \cite{Mark} into a set of
coupled quasilinear equations---called the
``representative'' set---whose {\it integrability condition} is exactly
the original nonlinear equation. We then tried the simplest possible
action functional of a single scalar field---including both combinations $G(x)$
and $\Lambda_{\pm}(x)$ as special cases---and coupled it to
gravity. It quite remarkably reconstructed the ``representative''
equation and produced the general 
solution of the differential equation of \cite{Mark}, modulo certain reality conditions. The result strongly suggests 
that the r\^ole of scalar fields in classical general relativity
deserves to be investigated further. 

Such an investigation should be partly related to the physical
interpretation of the action discussed here, as well as to some
geometrical understanding of the correlation between pure gravity---i.e.,
Markopoulou's equation---and scalar fields---the ``representative'' equation. In addition, the possibility of connecting the above
results with the more general problem of representing spacetime
diffeomorphisms in canonical gravity \cite{IsKu}\cite{Kuchar} is
certainly not excluded. It would also be interesting to know whether
these commuting combinations can arise from a canonical transformation
of the geometric data \cite{KuTime}, or even if they can be related
to the Ashtekar program
\cite{Asht} in quantum gravity, where {\it complex} combinations of the
geometric data are {\it also} allowed. We hope to be able to discuss these questions in the future.

\section{Acknowledgements.}
My thanks are due to C. Isham, C. Anastopoulos,
K. Garth, A. Kakou, F. Markopoulou, and K. Savvidou, for their many
suggestions and help.
I would also like to thank the ``Alexander S. Onassis Public Benefit
Foundation'' for their financial support.

\begin{appendix}
\section{Appendix}

We shall explicitly demonstrate equivalence between cases 
(\ref{eq:a}) and (\ref{eq:b}); the same argument applies when
comparing any two cases from the set (\ref{eq:a})--(\ref{eq:d}). We
respectively 
denote the set of functions obeying equations (\ref{eq:a}) and
(\ref{eq:b}), by $(\mtilde_1,\ltilde_1,R_1)$ and
$(\mtilde_2,\ltilde_2,R_2)$. For the two cases to be equivalent (in the sense
described in subsection 2.2) the following three conditions must be
satisfied: 
\begin{enumerate}
\item
$\mtilde_1$, $\ltilde_1$, $R_1$ should obey (\ref{eq:a})---the
functions $\mtilde_1$,
$\ltilde_1$ regarded as known---
\item
$\mtilde_2$, $\ltilde_2$, $R_2$ should obey (\ref{eq:b}) ~~~~~~~~~~~~~~~~~~and
\item
${\mit v}[\htilde,\ftilde]={\ltilde_1}^{\omega\over2}(\htilde-\mtilde_1+\rtilde_1)^{\omega\over2}=
{\ltilde_2}^{\omega\over2}(\htilde-\mtilde_2+\rtilde_2)^{\omega\over2}$.
\end{enumerate}

To avoid having to compare the differential equations of conditions 1 and 2, we do the following ``trick''.
We successively insert equations (\ref{eq:a}) and (\ref{eq:b}) into
equations  (\ref{eq:vf}) determining the partial
derivatives of ${\mit v}[\htilde,\ftilde]$, which by condition 3 must be the
same for both cases. As a result, we turn the differential equations 
(\ref{eq:a}) and (\ref{eq:b}) into the pair of  algebraic equations
\begin{equation}
{\mit v}_H={{\ltilde_1}^{\omega\over2}(\htilde-\mtilde_1+\rtilde_1)^{\omega\over2}\over\rtilde_1}=-
{({\ltilde_2})^{\omega\over2}(\htilde-\mtilde_2+\rtilde_2)^{\omega\over2}\over\rtilde_2}
\label{eq:iiia}
\end{equation}
\begin{equation}
{\mit v}_F={{\ltilde_1}^{\omega\over2}(\htilde-\mtilde_1+\rtilde_1)^{{\omega-2}\over2}\over\rtilde_1}={{\ltilde_2}^{\omega\over2}(\htilde-\mtilde_2+\rtilde_2)^{{\omega-2}\over2}\over\rtilde_2},
\label{eq:iiib}
\end{equation}
and---by  comparing condition 3 with equations (\ref{eq:iiia}) and
(\ref{eq:iiib})---we get the {\it consistent} solution (since the system is
over--determined for $\mtilde_2$, $\ltilde_2$ in terms of $\mtilde_1$,
$\ltilde_1$):  
\begin{equation}
\mtilde_2=2\htilde-\mtilde_1\qquad{\ltilde_2}^{\omega\over2}=({-\ltilde_1})^{\omega\over2}\qquad
\rtilde_2=-\rtilde_1,
\end{equation}
which proves equivalence.

For completeness, we write down the corresponding results from comparing
cases (\ref{eq:a}) with (\ref{eq:c}) and (\ref{eq:a}) with
(\ref{eq:d}); we still consider the pair $(\mtilde_1, \ltilde_1)$---that obeys 
(\ref{eq:a})---as given:
\begin{equation}
\mtilde_3=2\htilde-\mtilde_1,\qquad
({\ltilde_3})^{\omega\over2}={({-\ltilde_1})^{\omega\over2}(\htilde-\mtilde_1+\rtilde_1)\over
(\htilde-\mtilde_1-\rtilde_1)},\qquad
\rtilde_3=\rtilde_1,
\end{equation}
\begin{equation}
\mtilde_4=\mtilde_1,\qquad
({\ltilde_4})^{\omega\over2}={({-\ltilde_1})^{\omega\over2}(\htilde-\mtilde_1+\rtilde_1)\over
(\htilde-\mtilde_1-\rtilde_1)},\qquad
\rtilde_4=-\rtilde_1,
\end{equation}
where $(\mtilde_3,\ltilde_3)$ and $(\mtilde_4,\ltilde_4)$ respectively
satisfy equations (\ref{eq:c}) and (\ref{eq:d}).

\section{Appendix}

The exact relation between the quantities $\lambda[\htilde,\ftilde]$ and
$\mu[\htilde,\ftilde]$---used in the present paper---and the quantities 
$B(\alpha)$, $B'(\alpha)$ and $\alpha[\htilde,\ftilde]$---appearing in \cite{Mark}---is established.

By comparing equation (\ref{eq:aasol}) with the ``$\omega$--ansatz'' equation (\ref{eq:ansatz}), one immediately identifies 

\begin{equation}
\mu[\htilde,\ftilde]={1\over 2}B'(\alpha[\htilde,\ftilde])
\end{equation}
and
\begin{equation}
\lambda[\htilde,\ftilde]=\pm \exp
\Biggl(B(\alpha[\htilde,\ftilde])+{\omega\over2}{{B'(\alpha[\htilde,\ftilde])\over2}\over 
\sqrt{\bigl(\htilde - {1\over 2}
B'(\alpha[\htilde,\ftilde])\bigr)-\ftilde}}\Biggr)
\end{equation}
which, using (\ref{eq:asol}), can be written as 
\begin{equation}
\mu[\htilde,\ftilde]={1\over 2}
B'\biggl(-{\omega\over4R[\htilde,\ftilde]}\biggr)
\label{eq:mimi}
\end{equation}
and
\begin{equation}
\lambda[\htilde,\ftilde]=\pm \exp\Biggl[{\omega\over2}{\mu\over R[\htilde,\ftilde]}+B\biggl(-{\omega\over
4R[\htilde,\ftilde]}\biggr)\Biggr].
\label{eq:lala}
\end{equation}
$B'(-1/2R)$ denotes the total derivative of $B(-1/2R)$ with respect
 to $(-1/2R)$. 
Equations (\ref{eq:mimi}) and (\ref{eq:lala}) provide an implicit solution of the representative equation
(\ref{eq:a})---depending on one arbitrary function of one
variable. From the arguments in the previous sections one expects
this
solution to be the general solution of
(\ref{eq:a})---independently of the value of $\omega$. In the present
form---and using the weight--two ansatz map---the choice
$B(-1/2R)=0$ gives the Kucha\v{r}--Romano combination, while the choice $B(-1/2R)=-ln(-1/2R)$ produces the Brown--Kucha\v{r} one.

\end{appendix}


\vskip 2cm

\newpage

\begin{thebibliography}{99}

\bibitem{Isham} C.J.Isham, { Canonical Quantum Gravity and the Problem
of Time}, in {\it Integrable systems, Quantum Groups, and Quantum
Field Theories}, pages 157--288. Kluwer Academic Publishers, London, 1993.

\bibitem{ADM} R. Arnowitt, S. Deser, and C.W. Misner, in
{\it Gravitation: An Introduction to Current Research\/}, edited
by L. Witten (Wiley, New York,1962). 

\bibitem{KuDGR} K.V. Kucha\v{r},
in {\it Quantum Gravity 2: A second Oxford symposium}, edited by
C.J.~Isham, R.~Penrose and D.W.~Sciama  (Clarendon Press, Oxford,
1981). 

\bibitem{Dirac} P.A.M. Dirac, {\it  Lectures on Quantum Mechanics}
(Yeshiva University, New York, 1964).

\bibitem{BrKu} J.D. Brown and K.V. Kucha\v r, {\it Phys. Rev.\/} D
{\bf 51}, 5600 (1995).

\bibitem{KuRo}K.V. Kucha\v{r} and J.D. Romano, {\it Phys. Rev.\/} D
{\bf 51}, 5579 (1995). 

\bibitem{Mark}F.G. Markopoulou, {\it Imperial College Preprint, No 95--96/20}.

\bibitem{IsKu}C.J. Isham and K.V. Kucha\v{r}, {\it Ann. Phys.\/} (NY)
{\bf 164}, 288 (1985); {\bf 164}, 316 (1985). 

\bibitem{HoKuTe} S.A. Hojman, K.V. Kucha\v{r} and C. Teitelboim, {\it
Ann. Phys.} {\bf 96}, 88--135 (1976).

\bibitem{Kuchar} K.V. Kucha\v{r} and C.G. Torre, {\it Phys. Rev.\/} D
{\bf 43}, 419 (1991).

\bibitem{KuTime} K.V. Kucha\v{r}, {\it Matter time in quantum
gravity}, University of Utah preprint, UU--REL--92--12--10.

\bibitem{Asht} A. Ashtekar, {\it Phys. Rev. 36}, 1587, and {\it
Phys. Rev. Letts. 57}, 2244.


\end{thebibliography}
\end{document}